\documentstyle[epsfig]{elsart}
\begin{document}
\begin{frontmatter}
\title{Statistical Model Description of $K^+$ and $K^-$ Production 
between 1 - 10 $A\cdot$GeV}
%
\author[UCT]{J.~Cleymans},
\author[TUD]{H.~Oeschler},
\author[UW,GSI]{K.~Redlich}
\address[UCT]{Department  of  Physics,  University of Cape Town,\\
Rondebosch 7701, South Africa}
\address[TUD]{Institut f\"ur Kernphysik, 
Technische Universit\"at Darmstadt, D-64289~Darmstadt, Germany}
\address[UW]{Institute of Theoretical Physics, University of Wroc\l aw,
 Pl-45204 Wroc\l aw, Poland }
\address[GSI]{Gesellschaft f\"ur Schwerionenforchung, D-64291 Darmstadt, 
Germany}

\begin{abstract}
The excitation functions of $K^+$ and $K^-$ mesons in heavy ion collisions
are studied within a statistical model assuming chemical and thermal 
equilibrium with exact strangeness conservation.
At low incident energies the associate
production of kaons, i.e.~the production of a $K^+$ together with a hyperon and
the production of a $K^-$ together with a $K^+$, implies specific
features: different threshold energies and different dependences of 
$K^+$ and $K^-$ yields on baryon number density.
It is shown that the experimentally observed equality of the $K^+$ and $K^-$ 
rates
at energies $\sqrt{s} - \sqrt{s_{th}} \leq 0$ 
is due to a crossing of the two excitation functions.
Furthermore, the independence of the $K^+$ to $K^-$ ratio 
on the number of participating nucleons observed at 
1 and 10 $A\cdot$GeV is consistent with this model.
\end{abstract}
\end{frontmatter}

\newpage

Central heavy ion collisions at relativistic energies 
present an ideal tool to study
nuclear matter at high densities and  high temperatures.
However, these collisions are  complex and
in order to interpret the results, two strategies are commonly used:
(i) to describe the time evolution of the collisions
using transport models and (ii) to use
 a statistical concept
assuming thermal and chemical equilibrium and common freeze out
parameters for 
all particles.

In this Letter the second procedure, the statistical concept, will be followed.
Of special interest here is the production 
of  $K^+$ and $K^-$ below and above
the respective $NN$ thresholds. The experimental results have attracted much
interest as the measured $K^+$ to $K^-$ ratios in heavy ion collisions
differ strongly from the ratios obtained in $NN$ 
reactions~\cite{Barth,Laue}.
These findings have lead to the 
proposal that in heavy ion collisions
the ``effective masses'' of $K^+$ and $K^-$ are changed as predicted
for dense nuclear matter. 
The aim of this Letter is to discuss the $K^+$ and $K^-$ production
within a statistical model.
This model describes the condition at freeze out
using masses of free particles.

The production of strange particles has to respect strangeness
conservation. The attempts to describe the measured particle ratios
including strange hadrons at AGS and SPS using a strangeness fugacity
$\lambda_S$ is quite successful~\cite{Cley,PBM,Sollfrank,Heppe,Toneev}.
However, the usual grand-canonical treatment is not sufficient,
if the number of
strange particles is small~\cite{Hagedorn}. 
This requires exact strangeness conservation which
is done in the statistical model using the canonical formulation 
of strangeness conservation~\cite{CLE99}.
Consequently, the abundance of $K^+$ mesons is suppressed since
together with each $K^+$ also another strange particle,
e.g.~a $\Lambda$ hyperon is produced via $NN \rightarrow N \Lambda K^+$.
And for $K^-$ the corresponding channel is 
$NN \rightarrow N N K^- K^+$.
While the pion multiplicity per $A_{part}$ 
is approximately given by a simple Boltzmann factor 
(neglecting resonance contributions and isospin asymmetry),
\begin{equation}
\frac{M_{\pi}}{A_{part}} \sim \exp \left(-\frac{E_{\pi}}{T}\right),
\end{equation}
the multiplicity of positively charged kaons 
is given by 
\begin{equation}
\frac{M_{K^+}}{A_{part}} \sim \exp \left(-\frac{E_{K^+}}{T}\right)
\left[g_{\Lambda}V \int {d^3p\over (2\pi)^3}
\exp\left(-{{(E_{\Lambda}-\mu_B)}\over T}\right)\right],
\end{equation}
with the temperature $T$, the baryo-chemical potential $\mu_B$,
the degeneracy factors $g_i$, the volume  $V$ 
(see~\cite{CLE99}) and the  energies $E_i$ of the particles $i$ and 
integrating over momentum $p$.

The formula above, simplified for demonstration purpose,
neglects higher order terms in $V$~\cite{CLE99}, quantum statistics 
and other processes leading to the production of $K^+$.
The corresponding formula for $K^-$ production is similar, but does not
depend on $\mu_B$, 
\begin{equation}
\frac{M_{K^-}}{A_{part}} \sim \exp \left(-\frac{E_{K^-}}{T}\right)
\left[g_{K^+}V \int {d^3p\over (2\pi)^3}
\exp\left(-{E_{K^+}\over T}\right)\right].
\end{equation}

>From Eqs.~(1) - (3) it is obvious that the exact strangeness 
conservation implies
a reduction of $K^+$ and $K^-$ 
yields as compared to the values calculated
without exact strangeness conservation~\cite{CLE99}.

In addition, since the volume in Eqs.~(2) - (3) is proportional 
to the number of participants $A_{part}$, the $K^+$ and $K^-$ multiplicities 
are expected to rise 
(for low $T$ and small $V$) quadratically with $A_{part}$
while $M_{\pi}$ increases linearly with $A_{part}$.
These properties are in remarkable agreement with the experimental
observations~\cite{CLE99,Mang,HO98}.

The measured yields (or particle ratios)
can be described in this statistical concept by
lines  in the $T$ and $\mu_B$ plane.
All particle ratios measured around 1~$A\cdot$GeV 
(besides $\eta/\pi_0$) intersect
within the experimental errors reflecting  common values for
 $T$ and $\mu_B$ for 
all particles at freeze out~\cite{CLE99}.
Surprisingly, even the measured $K^+/K^-$ ratio
fits into this picture and the calculated ratio does not depend 
on the choice
of the volume  $V$. However, the $A_{part}$ dependence enters 
in the ratios of strange to non-strange particles, e.g.~in $K^+/\pi^+$.

\begin{figure}
\mbox{\epsfig{width=10cm,file=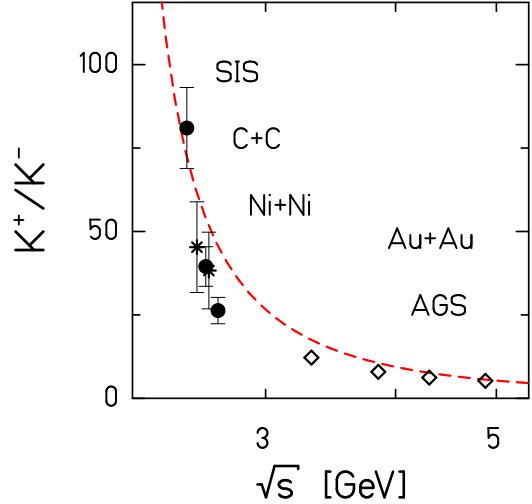}}
\caption{{\small Measured $K^+/K^-$ ratios ($\bullet$ for C+C collisions,
\cite{Laue}, $\ast$ for Ni+Ni \cite{Barth,FRS},
$\diamond$ for Au+Au \cite{Dunlop}) as a function of $\sqrt{s}$
together with a description within the statistical model given as dashed line.
}}
\label{KPKM_ratio}
\end{figure}

Figure~\ref{KPKM_ratio} shows the $K^+/K^-$
ratios measured by the KaoS Collaboration
\cite{Barth,Laue}, by the FRS Group \cite{FRS} 
and by the E866/E917 Collaboration at AGS
\cite{Dunlop} as a function of $\sqrt{s}$.
To obtain the theoretical results, shown as dashed line, we
start from the universal freeze-out curve suggested in \cite{CLEY98}.
Together with the measured systematics for the pion multiplicities,
relations for $T$ and $\mu_B$ as functions of
$\sqrt{s}$ are obtained \cite{CLEY99b}.
Within this approach, the  $K^+/K^-$ ratios are 
given as a dashed line in Fig.~\ref{KPKM_ratio}.
The observed rise towards low incident energies reflects the fact
that the two kaon species have different threshold energies due to
their associate production.

The $K^+/K^-$ ratios measured in heavy ion collisions
by the KaoS Collaboration show that
the $K^-$ yield compared to the $K^+$ cross section
is much higher than expected from $NN$ collisions \cite{Barth,Laue}.
This is especially evident, if the kaon multiplicities are
plotted as a function of $\sqrt{s} - \sqrt{s_{th}}$ where
$\sqrt{s_{th}} - 2m_N$ is the energy needed to produce the corresponding
particles taking into account the mass of the produced partners
($\sqrt{s_{th}(K^+)} - 2m_N$ = 0.67 GeV, 
$\sqrt{s_{th}(K^-)} - 2 m_N$ = 0.987 GeV).
The measured $K^+$  and $K^-$ yields in heavy ion
collisions are about equal for $\sqrt{s} - \sqrt{s_{th}} \leq 0$ 
while the $K^+$ yield in $NN$ collision exceeds
the $K^-$ yields by a factor of 10 -- 100 close to threshold.

\begin{figure}
\mbox{\epsfig{width=10.5cm,
file=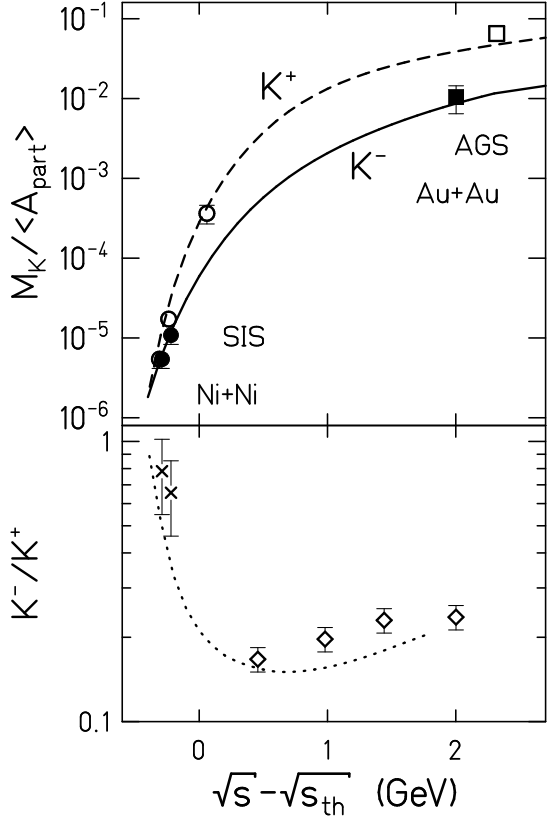}}
\caption{{\small Upper part: Calculated $M_K/A_{part}$ 
ratios
in the statistical model as a function of $\sqrt{s} - \sqrt{s_{th}}$.
The dashed (solid) line refers to $K^+$ ($K^-$).
Open (full) symbols represent measured $K^+$ ($K^-$) multiplicities.
Lower part: $K^+/K^-$ ratios from the excitation functions above together
with results from Ni+Ni collisions ($\times$) at SIS energies and Au+Au 
at AGS energies ($\diamond$).}}
\label{KP_KM_sthr_therm}
\end{figure}

In Fig.~\ref{KP_KM_sthr_therm} we show in the upper part the
multiplicities of $K^+$ and $K^-$ divided by $A_{part}$ as a function of 
$\sqrt{s} - \sqrt{s_{th}}$ over a large energy range from SIS up to AGS. 
The full and dashed lines refer to the statistical 
model results for $K^-$ and $K^+$ respectively.
At values of $\sqrt{s} - \sqrt{s_{th}}$ less than zero
the two excitation functions cross.
They
differ at AGS energies by a factor of five which is in good agreement
with the result for central collisions of Au+Au at 
10.8 $A\cdot$GeV~\cite{Ahle98}.
The model calculations depend on the choice of the system,
here Ni+Ni collisions.
At SIS energies, only inclusive measurements for Ni+Ni are available.
The values for $K^+$ are from Ref.~\cite{Barth}. 
The results for $K^-$
are from Ref.~\cite{FRS} and corrected for the angular distribution~\cite{Marc}.
$A_{part}$ is chosen as $A$
which is based on estimates from the mean $A_{part}$ for $K$ production
as kaons originate more from central collisions.
At AGS energies the choice of the system has little 
influence which allows to plot the results for Au+Au collisions 
at 10.8 $A\cdot$GeV as well.
This figure evidences that the similarity of the $K^+$ and $K^-$ yield
observed around 1 -- 2 $A\cdot$GeV arises from the difference in
the rise of the two excitation functions.
This difference can be understood by the approximate formulae given in 
Eqs.~(2) and (3).
The density of $K^+$ contains  the term $E_{\Lambda}-\mu_B$ while the
$K^-$ density has $E_{K^+}$ in the exponent. As these two values are 
different, the
excitation functions, i.e.~their variation with $T$, exhibit different slopes.

\begin{figure}
\mbox{\epsfig{width=8.5cm,file=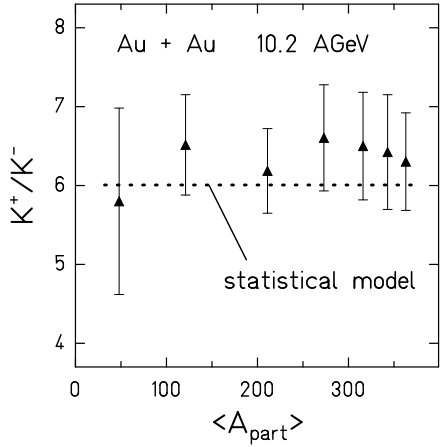}}
\caption{{\small Ratio of $K^+$ to $K^-$  as a function
of the number of participants $A_{part}$ from \cite{Ahle99}
together with the statistical model (dashed line) evidencing the
independence of $A_{part}$. }}
\label{Apart_ags}
\end{figure}

Furthermore, Eqs.~(2) and (3) evidence that for a low 
temperature $T$ and a small volume $V$
the dependence of the $K^+$ and $K^-$ 
multiplicity on $A_{part}$ is quadratic
which is in very good agreement with 
data~\cite{CLE99,Mang,HO98}.
Small variations from the quadratic dependence can occur 
due to a change of $T$ and $\mu_B$ with $A_{part}$~\cite{CLE99}.
It is interesting to note that also hydrodynamical models
predict a variation of $K^+$ and $K^-$ with $A_{part}^2$~\cite{Russkikh,Kolo}. 
Transport models, on the other hand, show an increase
with $A_{part}^{\alpha}$ where $\alpha$ is approximately  
1.4 - 1.6~\cite{Hartnack}.
In these models
the kaons are produced in multi-step processes
 which are more likely in central
collisions where the density is higher.
As already mentioned,
the statistical model predicts that the variation of the $K^+$ and 
of the $K^-$ yields
with $A_{part}$ are equal. Hence, for a given collision
the $K^+/K^-$ ratio is expected not to vary with centrality.
Indeed, this is in accordance with the data for Au+Au collisions at
10.2~$A\cdot$GeV~\cite{Ahle99}.
Figure~\ref{Apart_ags} shows the results together with the
prediction of the statistical model.
It has even been observed at SIS energies for Ni+Ni collisions at 
1.93~$A\cdot$GeV~\cite{Marc}.
This is remarkable as the
$K^+$ production is above and the $K^-$ production below their respective $NN$ 
thresholds.

In summary,
the statistical model using exact strangeness conservation
is able to describe most of the measured particle ratios 
from SIS up to SPS energies.
Within this framework the equality of $K^+$ and $K^-$ multiplicities
as a function of $\sqrt{s} - \sqrt{s_{th}}$ is  a consequence of two
excitation functions with different slopes crossing 
at values $\sqrt{s} - \sqrt{s_{th}}$ below zero.
This model is also able to describe the dependence of the kaon 
yields on $A_{part}$
being quadratic around 1~$A\cdot$GeV. This effect fades away 
with increasing
incident energy.
The $K^+/K^-$ ratio is predicted in the considered model as being  
independent of $A_{part}$ and
this is, indeed, observed from SIS up to AGS energies.

The statistical model presented here uses a unique freeze out for all particles.
Detailed experimental studies on pion production
show evidence for a time evolution of the
pion emission with high-energy pions being emitted earlier~\cite{AW99}.
Such effects, however, are not visible on the level of total particle
multiplicities since these involve integrals over the whole phase space.
Despite  the apparent success of the statistical model of
particle production under the assumption of thermal and
chemical equilibration and using masses of free particles, 
the present understanding of hadronic
interactions contradicts chemical equilibrium
for strange particles~\cite{Elena,Bravina} 
This discrepancy seem
 to put into question our present understanding
of interactions at the high densities reached in heavy ion collisions.
Indeed, already at and above twice nuclear matter densities,
nucleons are hardly ``free'' individual particles. 
This ``in-medium'' effect clearly deserves
further studies.

K.R. acknowledges the partial support of the State Committee for Scientific
Research (KBN).


\end{document}